\documentclass[
aps,%
12pt,%
final,%
notitlepage,%
oneside,%
onecolumn,%
nobibnotes,%
nofootinbib,% In the current version of REVTeX, when this option is on,
superscriptaddress,%
noshowpacs,%
centertags]%
{revtex4}

\begin{document}
\selectlanguage{english}
\title{ STRUCTURE OF ODD Ge ISOTOPES WITH $40 < N < 50$}
\author{\bf{\firstname{P.} \surname{C.} Srivastava$^{1)*}$}}  

\footnote{$^{1)}$Instituto de Ciencias Nucleares, Universidad Nacional Aut\'onoma de M\'exico,  M\'exico.}
\author{\bf{\firstname{M.} \surname{J.} Ermamatov$^{1),2)}$}} 
\footnote{$^{2)}$Institute of Nuclear Physics, Ulughbek, Tashkent, Uzbekistan.\\
$^*$Email: praveen.srivastava@nucleares.unam.mx}
\begin{abstract}   
%\vspace{1.0cm}     

We have interpreted recently measured experimental data of $^{77}$Ge, and
also for $^{73,75,79,81}$Ge isotopes in terms of state-of-the-art shell model
calculations. Excitation energies, $B(2)$ values, quadrupole moments 
and magnetic moments are compared with experimental data when available.
 The calculations have been performed with the recently
derived interactions, namely with JUN45 and jj44b for 
${f_{5/2}pg_{9/2}}$ space. We have also performed calculation 
for ${fpg_{9/2}}$ valence space using an ${fpg}$ effective interaction with $^{48}$Ca core and
imposing a truncation to study the importance of the proton excitations 
across the $Z=28$ shell in this region. The predicted results of jj44b interaction are in good
agreement with experimental data.

\end{abstract}

%\pacs{21.60.Cs, 27.50.+e} 
\maketitle

%\newpage
\section{Introduction}
\label{s_intro} 

 Nuclear structure study  in $40 \leq N \leq 50$ region is the topic of current research for the investigation
of single-particle versus collective phenomena. For $fp$ shell nuclei the collective phenomena was
measured in different laboratories worldwide. In the nuclei where 
$\pi0f_{7/2}$ shell is not completely filled and $N \sim 40$, 
deformation appears due to interaction of 
excited neutrons in $sdg$ shell with protons in $fp$ shell.
The experimental indication of a new region of deformation for $fp$ shell nuclei has been reported for Fe \cite{Rother11}, Cr~\cite{Aoi09} and
Co~\cite{Recchia12} isotopes. The importance of  the inclusion of intruder orbitals from $sdg$
shell in the model space for $fp$ shell nuclei is reported in the literature~\cite{Kaneko08,Sri09,Sria,Sriaa,Srie,Lenzi09,Sun12}. 

\begin{figure}
\begin{center}
%\resizebox{90mm}{!}{\includegraphics{Fig1.eps}}
\includegraphics[width=16.4cm]{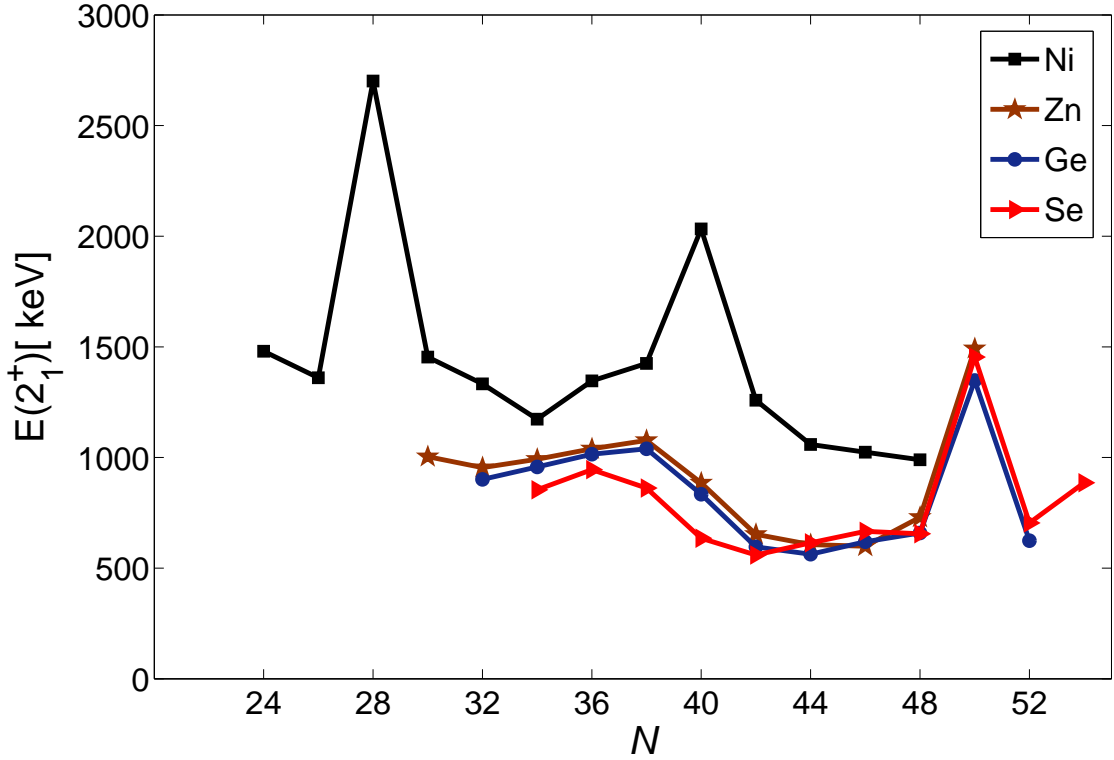}
\caption{ Experimental $E(2_1^+)$ for even--even nuclei.} 
\label{exp}
\end{center}
\end{figure}

The nuclei around Ni region, particularly, Ga and Ge isotopes are interesting from both
experimental and theoretical point of view. Sudden structural changes between $N=40$ and $N=50$ for Ga isotopes have been observed at \textsc{isolde, 
cern} \cite{Cheal10}. The even Ge isotopes attracted much experimental and theoretical 
interest due to a shape transition in the vicinity of $N=40$~\cite{Rodal05,Zamick11,Sri12con}.  
The low-lying excitation energies of Ge isotopes  show unexpected behavior: as we move from $^{70}$Ge to $^{76}$Ge the  
$E(0_2^+)$ for $^{72}$Ge is lower
than $E(2_1^+)$, it again start rising from $^{74}$Ge onwards.  In the pioneering work of Padilla-Rodal et al. \cite{Rodal05}  
the evolution of collectivity in Ge isotopes with $B(E2)$ measurements have been reported. It was shown 
that the $N=40$ shell closure is collapsed in $^{72}$Ge. However, the $N=50$ shell closure is persistent in Ge isotopes. 
Collectivity at $N=50$
for the $^{82}$Ge and $^{84}$Se was measured using the intermediate-energy Coulomb excitation 
\cite{Gade10}. In Fig.~\ref{exp}, we show the experimental $E(2_1^+)$ for even-even nuclei.  
The  rapid decrease of $E(2_1^+$) for the Ge and Se isotopes around $N=40$  reveals the collapse of this shell closure.

In the present paper we consider neutron-rich odd Ge isotopes. The shell model calculation in $f_{5/2}pg_{9/2}$ space for Ge isotopes is reported in the
literature using pairing 
plus quadrupole-quadrupole interactions~\cite{yosinaga} and with JUN45 interaction for $^{73}$Ge and $^{75}$Ge in \cite{Honma09}. In this work we report shell model calculations in $f_{5/2}pg_{9/2}$  space with the recently devised effective 
interactions, we also first time report shell model results for odd Ge isotopes including $f_{7/2}$ orbital in the model space to study importance of 
proton excitations  across $Z=28$ shell as suggested in  \cite{Cheal10}.  The electromagnetic moments of Ga isotopes were explained in our 
recent investigation~\cite{pcs_ga} by including $f_{7/2}$ orbital in the $f_{5/2}pg_{9/2}$ model space.

Section~2 gives details of the shell model (SM) calculations. We will discuss in this section the  model space and the effective interactions used in the investigation.
Section~3 includes results on the spectra of $^{73-81}$Ge odd isotopes  and configuration mixing in these nuclei.
In Section~4, SM calculations on $E2$ transition probabilities, quadrupole moments and magnetic moments are presented. 
Finally, concluding remarks are given in Section~5. 

\section {\label{sec2}Details of Model Spaces and Interactions}

In the present study we have performed calculations in two different shell-model spaces.
The first set of calculations have been performed 
with the two recently derived effective shell model interactions, JUN45
and jj44b, that have been proposed for the 1$p_{3/2}$,
0$f_{5/2}$, 1$p_{1/2}$ and 0$g_{9/2}$ single-particle
orbits.  The JUN45, which was recently developed by  Honma {\it et al.} \cite{Honma09}, is a realistic interaction based on the Bonn-C potential fitting by 400 experimental binding and excitation energy data with mass
numbers $A = $ 63--96.  The jj44b interaction was developed by Brown and Lisetskiy \cite{brown} by fitting 600 binding energies and excitation energies with $Z = $ 28--30
and $N = $ 48--50. Instead of 45 as in JUN45, here 30 linear combinations of good $J$--$T$ two-body
matrix elements (TBME) varied, with the rms deviation of about 250 keV from experiment.
The shell model results based on jj44b interaction have recently been reported in the literature
\cite{kum12,pcs_ga,Cheal10,saha12}. The 
second set of calculations have been performed in the  $f \,p \,g_{9/2}$ valence space,
we use a  $^{48}$Ca core, where eight neutrons are frozen in the $\nu f_{7/2}$
orbital. This interaction was reported  by Sorlin {\it et al}.
\cite{Sorlin02}. As here the dimensions of the matrices become very large, a
truncation has been imposed. We allowed up to a total of four particle
excitations from the $f_{7/2}$ orbital to the upper $fp$ orbitals for protons and from the upper $fp$
orbitals to the $g_{9/2}$ orbital for neutrons.
The $fpg$ interaction was built
using $fp$ two-body matrix elements (TBME) from
 \cite{Pov01} and $rg$ TBME ($p_{3/2}$, $f_{5/2}$, $p_{1/2}$ and $g_{9/2}$ orbits) from \cite{Nowacki96}. For
the common active orbitals in these subspaces, matrix elements
were taken from  \cite{Nowacki96}. As the latter
interaction ($rg$)  was defined for a $^{56}$Ni core, a scaling
factor of $A^{-1/3}$ was applied to take into account the
change of radius between the $^{40}$Ca and $^{56}$Ni cores. The
remaining $f_{7/2} g_{9/2}$  TBME are taken from  \cite{Kahana69}.
 For the JUN45 and jj44b interactions the single-particle energies are based on those of $^{57}$Ni. In the second set of calculations for the $fpg$ interaction with 
the $^{40}$Ca core the single-particle energies are based on those of $^{41}$Ca.  In the present calculations the JUN45 and jj44b interactions estimate binding energies of nuclei, however  $fpg$ interaction gives relative energies with respect to $f_{7/2}$ orbital not binding energies. For the both set of calculations the single-particle energies are the same for protons and neutrons. 

\begin{figure}
\begin{center}
%\resizebox{90mm}{!}{\includegraphics{monopole.eps}}
\includegraphics[width=14.4cm]{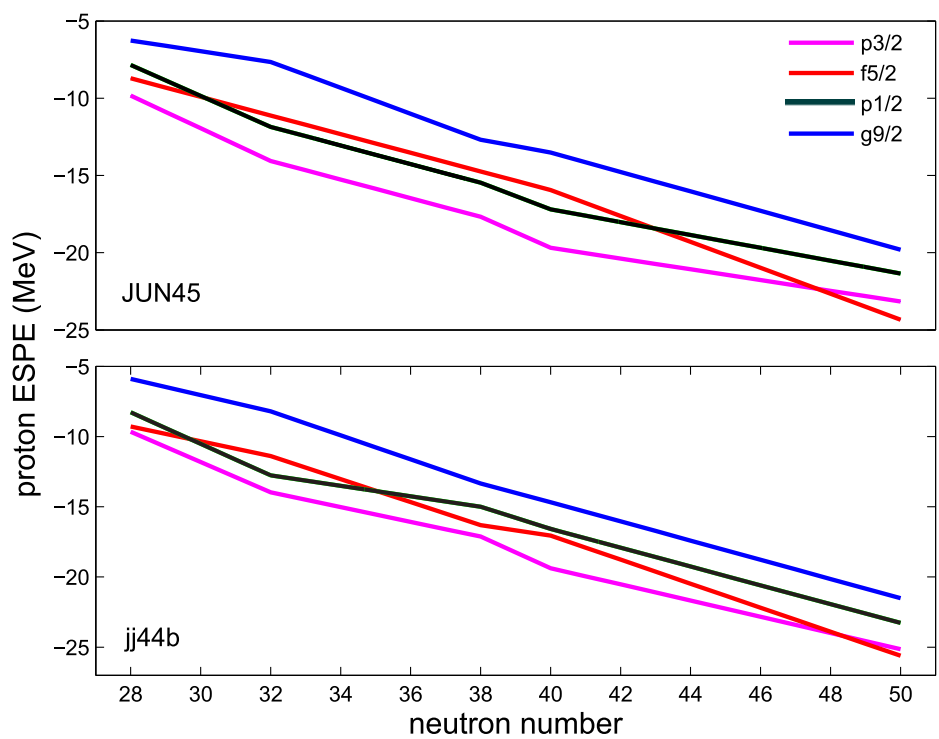}
\caption{Effective single--particle energies of proton orbits for Cu 
isotopes for JUN45 and jj44b interaction.} 
\label{eff}
\end{center}
\end{figure}

The single-particle energies for both protons and neutrons for the three interactions
are given in Table~1. In Fig.~\ref{eff}, we compare the effective single-particle energy
of  the proton  orbit for Cu isotopes in JUN45 and jj44b
interactions.  Both interactions show a rapid decrease
in $f_{5/2}$ proton single-particle energy relative to
$p_{3/2}$ as the neutrons start filling in $g_{9/2}$ orbit and
it become lower than $p_{3/2}$  for $N>48$.
Dimensions of the matrices for odd $^{73-81}$Ge  isotopes in the m-scheme for 
$f_{5/2}pg_{9/2}$  and $fpg_{9/2}$ spaces are shown in the Table~2. In case of $^{73}$Ge  
computing time $\sim$ 15 days for both parity. Obviously, maximal dimension $\sim 10^{8}$ is reached in the case of $^{73}$Ge when using
$f_{5/2}pg_{9/2}$ space with $^{56}$Ni core since neutron number is furthest from the closed shell for this nucleus among the Ge isotopes considered
in this work.

All calculations in the present
paper are carried out at \textsc{dgctic-unam} computational facility KanBalam
using the shell model
code \textsc{antoine}~\cite{Antoine}.

\section{\label{sec3}Spectra}
 
As is discussed in Section~\ref{s_intro}, for  $N<40$ and $N>40$ nuclei have different structural properties. 
This can be interpreted as a transition from
the spherical (or oblate) to prolate shape. It may also
point a coexistence of these phases. Such a change can be
seen in various experimental observables like nucleon transfer
cross sections, $B(E2)$ values and their ratios for
low-lying states~\cite{Honma09}.
The measured systematics show
the narrowing of
the $N = 40$ shell gap toward $Z = 32$~\cite{hak08}, while the persistence
of the $N = 50$ shell closure is suggested in $^{80}$Ge based on the
new $B(E2)$ data~\cite{iwa08}. The shell model structure of Ge isotopes was  discussed in~\cite{Honma09} using JUN45 interaction.
We shall start from the $^{73}$Ge which has 41 neutrons and is closest neighbor of $^{72}$Ge which has particular structure. 
Structural change can be observed by increasing neutron number from 41 ($^{73}$Ge) to 49 ($^{81}$Ge). The results for the three interactions used in the 
calculations
are presented with respect to the experiment.

\subsection{$^{73}$Ge}
 
\begin{figure}
%\setcaptionmargin{5mm}
%\onelinecaptionsfalse
\includegraphics[width=16.4cm]{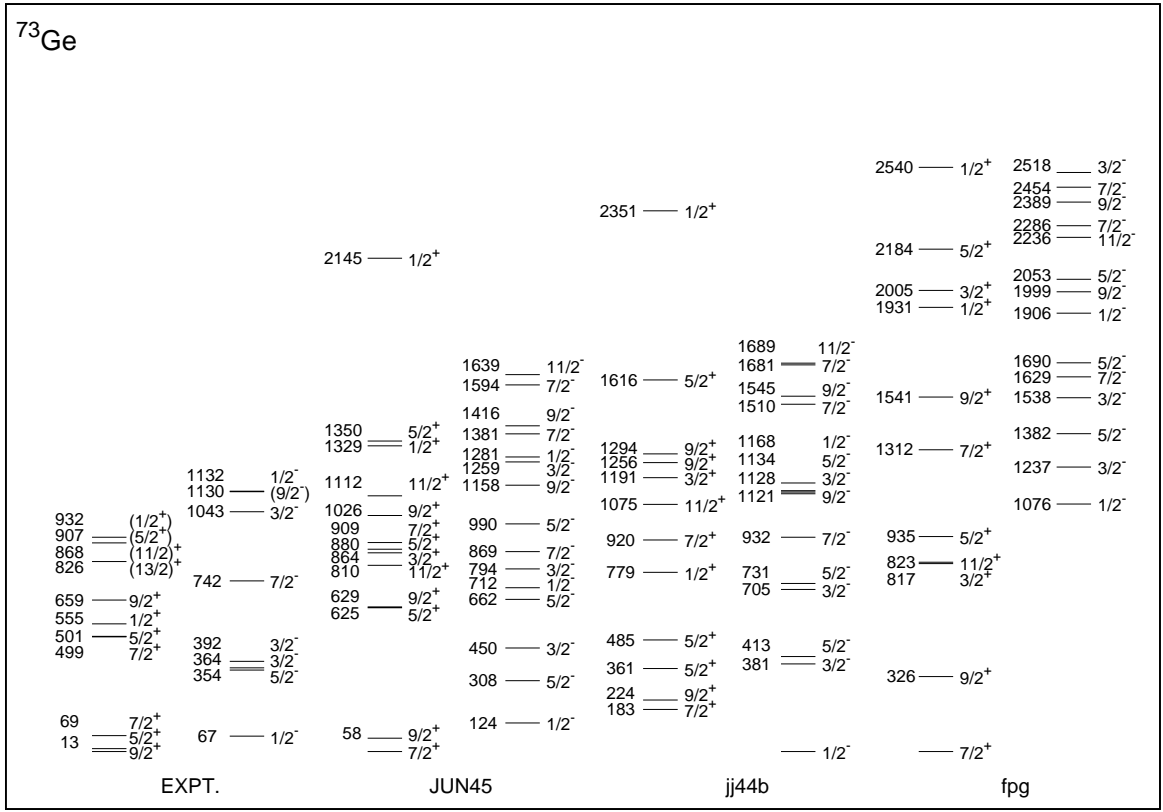}
%\captionstyle{normal} 
\caption{
Experimental data  for $^{73}$Ge \cite{nndc} compared
with the results of large-scale shell-model calculations using three different effective interactions}
\label{f_ge73}
\end{figure}

Comparison of the calculated values of the energy levels of $^{73}$Ge with the experimental data is shown in  Fig.~\ref{f_ge73}. 
All the three interactions fail to predict the ground state correctly. Difference in the triplet of levels mentioned in~\cite{Honma09} is 
the same in $fpg$ calculation too, i. e. $5/2_1^+$ is  high
in the calculation as compared to $7/2_1^+$ and $9/2_1^+$ levels, while in the experiment these three levels are arranged very close.
 Also $1/2^+_1$ is too high 
as compared to the experiment with jj44b and $fpg$ interactions. This was the case also for  the $^{69,71}$Ge~\cite{Honma09}. It was
supposed in this work that this could be because of
the  closed  $fp$ neutron shell and $d_{5/2}$ might play an important role.  However, it can 
be noted that $5/2^+_1$ and $1/2^+_1$ levels are very close to the other two levels in jj44b calculation.   Lower negative-parity levels 
are described well by JUN45
and jj44b calculations. Negative-parity levels are too high in $fpg$ calculation, though their arrangement looks like as in the other two calculations.
For the $9/2_1^+$ level the JUN45 and jj44 interactions predict $\nu(g_{9/2}^{3})$ (probability $\sim 12\%$)  and $\nu(g_{9/2}^{5})$ 
(probability $\sim 5\%$) configurations, respectively.

\subsection{$^{75}$Ge}
\begin{figure}
\includegraphics[width=16.4cm]{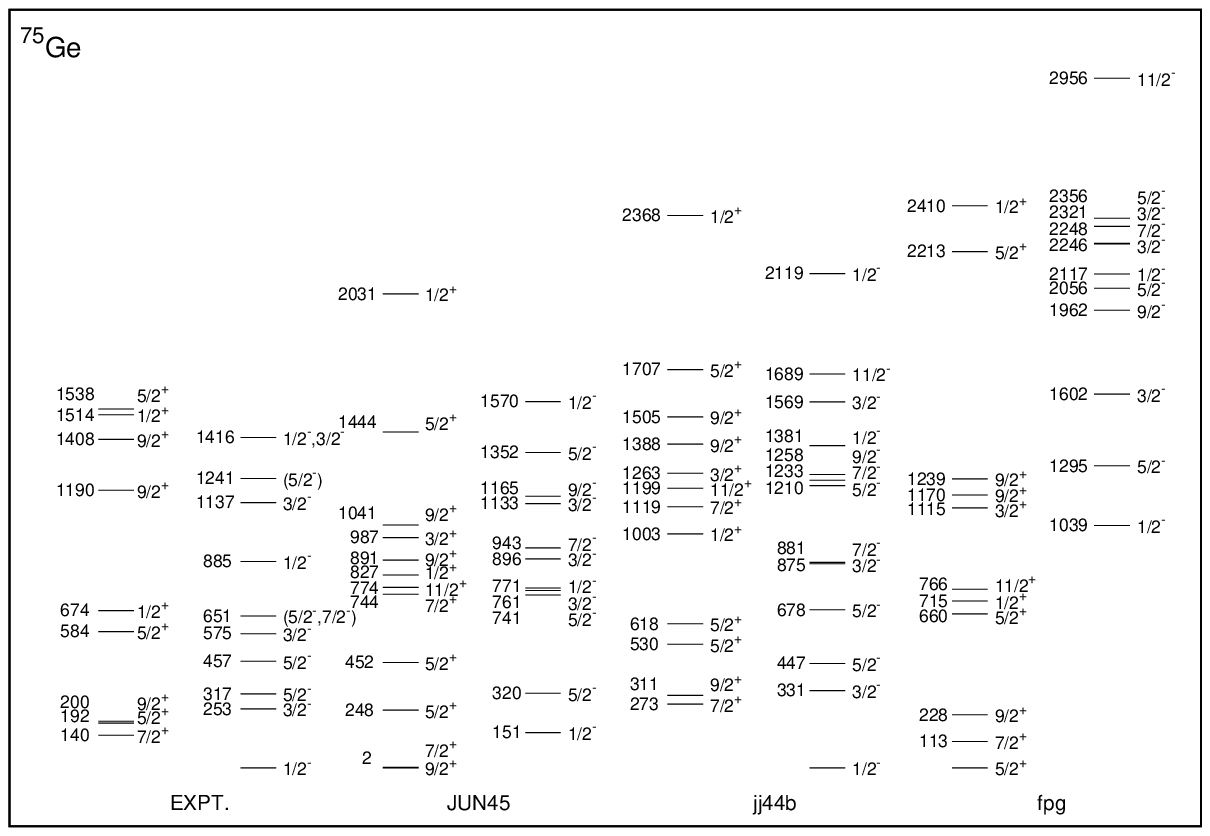}
\caption{
 The same as in Fig. 3, but for $^{75}$Ge. }
\label{f_ge75}
\end{figure}

Figure~\ref{f_ge75} shows the experimental and calculated  positive and negative parity levels of $^{75}$Ge using JUN45, jj44b and $fpg$ interactions.
 As is seen from the Fig.~\ref{f_ge75} that only jj44b gives correct ground state and location of positive and negative parity levels with respect 
 to the experiment.  For the negative-parity levels jj44b predicts correct spin sequence up to 
$3/2^-_2$. The jj44b interaction predicts $7/2^-$ for the experimentally uncertain $(5/2^-,7/2^-)$ level. 
With this interaction the experimental levels
$1/2_2^-$ at 885 keV, $3/2^-_3$ at 1137 keV, $(5/2^-_3)$ at 1241 keV and ($1/2^-,3/2^-$) at 1416 keV are predicted at 1381, 1569, 1210 and 2119 keV, 
respectively. 
   
The first positive parity $7/2^+$ level at 140 keV is predicted at 273 keV by jj44b calculation. The experimentally observed  $5/2^+_1$ is very close to $9/2^+_1$ level, while in the jj44b calculation $5/2^+_1$ is higher and well separated from $9/2^+_1$. 
 The next following levels are $5/2^+_2$ and $1/2^+_1$ both in the experiment and in calculation. Again in the calculation spacing between 
 these levels is larger than in the experiment.   There are $7/2^+_2$, $11/2^+_1$ and $3/2^+_1$ levels, between $1/2^+_1$ and 
 $9/2^+_2$ in the jj44b calculation. The experimental levels $1/2^+_2$ and $5/2^+_3$ at 1514 and 1538 keV are predicted at 2368 and 1707 keV, 
 respectively with jj44b interaction. 

The JUN45 and $fpg$  predict positive-parity - $9/2^+$ and $5/2^+$ g.s., respectively. The positive-parity levels are located lower and 
the negative-parity levels are located higher than the experimental levels. For the $9/2_1^+$ the JUN45 and jj44b interactions  predict 
$\nu(g_{9/2}^{5})$ ($\sim 15\%$) and $\nu(g_{9/2}^{5})$ ($\sim 18\%$) configurations, respectively. 
The $1/2_1^-$ has configuration $\nu(p_{1/2}^{-1})$
 with  JUN45 (20\%) and jj44b (21\%). The $fpg$ interaction predicts
 $\nu(g_{9/2}^{5})$ ($\sim 8\%$) configuration for the  $9/2_1^+$.

\subsection{$^{77}$Ge}

\begin{figure}
\includegraphics[width=16.4cm]{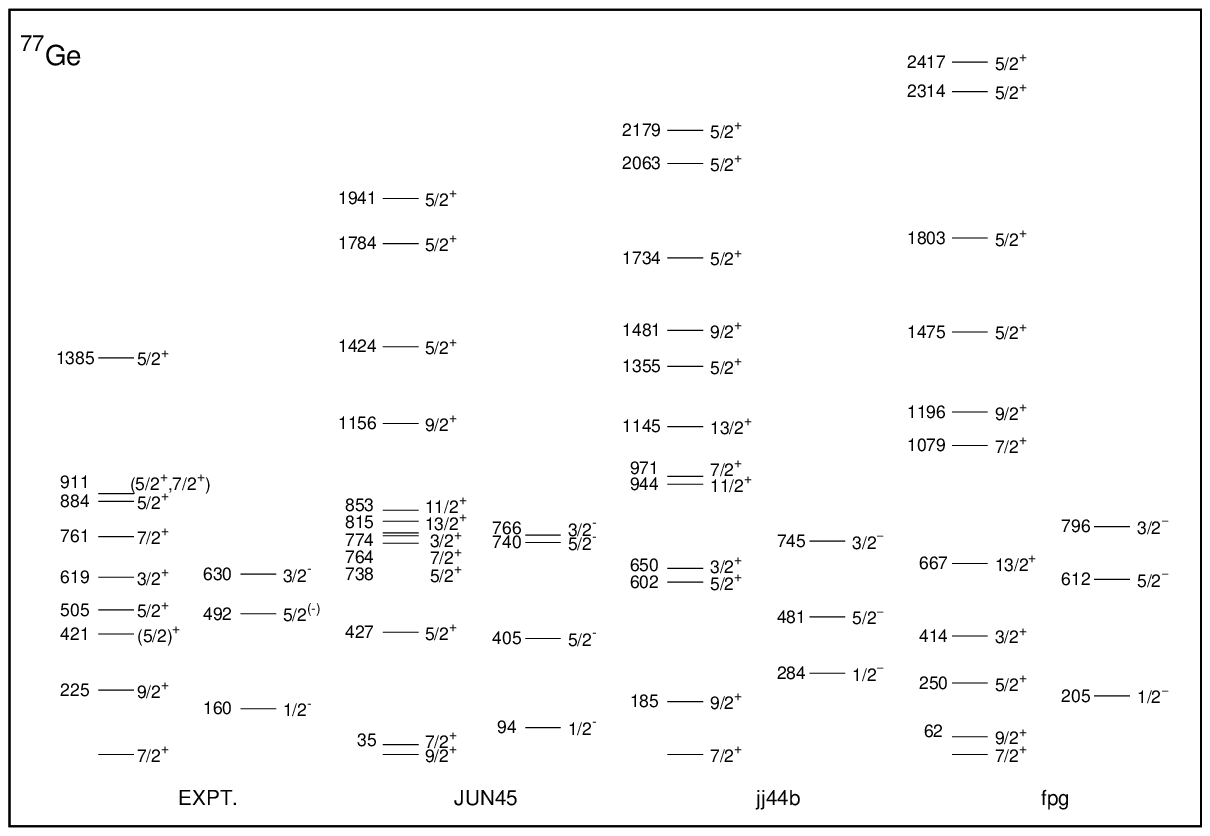}
\caption{
 Experimental data \cite{kay09} for $^{77}$Ge compared
with the results of large-scale shell-model calculations using three different effective interactions}
\label{f_ge77}
\end{figure}

Comparison of the calculated positive and negative parity levels with the recent experimental data at ATLAS facility~\cite{kay09} 
are shown in Fig.~\ref{f_ge77}. The $7/2^+$ ground state is now correctly predicted
not only by jj44b, but also with $fpg$ interaction. The $9/2_1^+$ is much closer
 to the ground state in $fpg$ calculation than in the experiment and jj44b calculation. 
The JUN45 calculation still gives $9/2^+$ ground state while the $7/2^+$ comes close to the ground state, however it will flip in $^{79}$Ge. 
The $5/2^+_1$ level is 
located only 6 keV higher than in the experiment in JUN45 calculation. The jj44b calculation result for this level is higher by 181 keV, while $fpg$ 
calculation for this level predict 161 keV less than in the experiment.  Next experimental level is $5/2_2^+$. For this level the result of  
JUN45 is 233 keV higher and is in the same sequence as in the experiment. In both jj44b and $fpg$ calculations this level is located very
high as compared to experiment and ordering of the levels is different from the experiment. The $3/2^+_1$ experimental level at 619 keV is predicted by 
jj44b calculation 31 keV higher than in the experiment. This level is predicted by JUN45 is 155 keV higher, while in $fpg$ it is 215 keV lower than in the experiment. The experimental level $7/2^+_2$ 
at 761 keV is predicted with difference of only 3 keV by JUN45. The jj44b and $fpg$ predictions for this level are 971 and 1079 keV, respectively.
 In all the calculations $5/2^+_3$, $5/2^+_4$ and $5/2^+_5$ levels are much higher than in the experiment. 

Three negative-parity levels are measured in the experiment. The $fpg$ calculation gives the closest value for the $1/2^-_1$ level. 
The experimental $5/2^-_1$ and $3/2^-_1$ are better predicted by jj44b calculation.  For the $9/2_1^+$, the JUN45 
and jj44b
interactions  predict 
$\nu(g_{9/2}^{5})$ ($\sim 20\%$)
 and $\nu(g_{9/2}^{7})$ ($\sim 19\%$) configurations, respectively. The $1/2_1^-$ has configuration $\nu(p_{1/2}^{-1})$ with probability 29\%
 and 14\% for JUN45 and jj44b interactions, respectively. The $fpg$ interaction predicts $\nu(g_{9/2}^{7})$ ($\sim 6\%$)
 configuration.

\subsection{$^{79}$Ge}
\begin{figure}
%\setcaptionmargin{5mm}
%\onelinecaptionsfalse
\includegraphics[width=16.4cm]{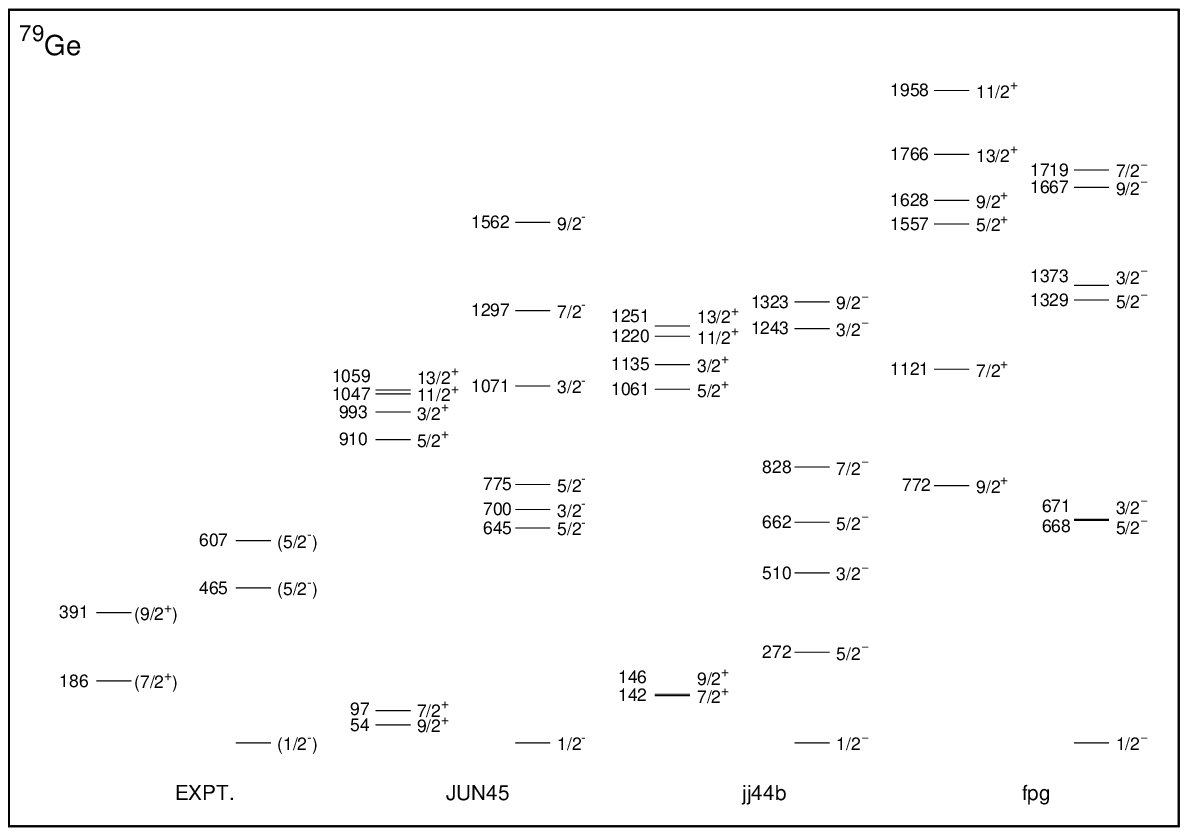}
%\captionstyle{normal} 
\caption{
 The same as in Fig. 3, but for $^{79}$Ge.}
\label{f_ge79}
\end{figure}

For this isotope few positive and negative-parity levels are available.
As is seen from Fig.~\ref{f_ge79} all the three interactions give correct ground state when reaching to this nucleus. The location of
$5/2^-_1$ and $5/2^-_2$  levels with respect to experimental ones  is good in JUN45. In jj44b these levels are lower than JUN45. The
$fpg$ interaction predicts $5/2^-_1$ similar to JUN45, but $3/2^-_1$ is higher to this level by only 3 keV, while for jj44b they are lower than JUN45 and $fpg$.  

The JUN45 and $fpg$ predict  reverse sequence of measured $(7/2^+)$ and $(9/2^+)$ levels. They are lower in JUN45 calculation and higher in 
$fpg$  calculation as compared to the experiment. The sequence of these levels are the same as in the experiment in jj44b calculation, however they 
are located very close to each other. For the $9/2_1^+$ the JUN45 and jj44b interactions  predict $\nu(g_{9/2}^{7})$ configuration with probability 34\%
and 35\%, respectively. For the $1/2_1^-$ the JUN45 and jj44b interactions  predict $\nu(p_{1/2}^{-1})$ configuration with probability 27\%
and 28\%, respectively.

\subsection{$^{81}$Ge}

\begin{figure}
%\setcaptionmargin{5mm}
%\onelinecaptionsfalse
\includegraphics[width=16.4cm]{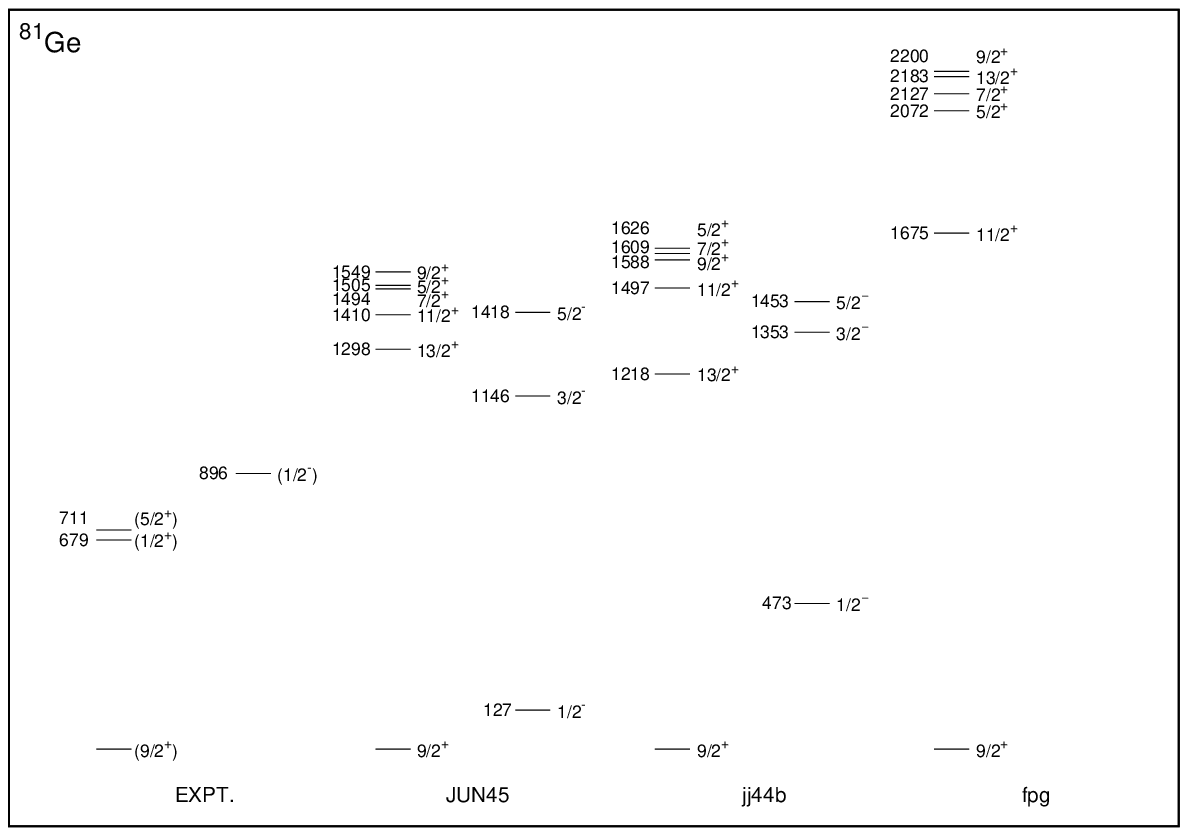}
%\captionstyle{normal} 
\caption{
 The same as in Fig. 3, but for $^{81}$Ge.}
\label{f_ge81}
\end{figure}

Three uncertain positive-parity and one negative-parity levels are available for this isotope. Again all calculations predict the
same parity and spin as in the experiment for the ground state. The other experimental positive-parity levels are predicted  higher than in the experiment by all the calculations. The only measured $1/2^-$ negative-parity level is low in the JUN45 and jj44b. In $fpg$ calculation negative-parity levels are very high ($>$ 4 MeV), thus we have not included them in Fig.~\ref{f_ge81}.
 
 Structure of the wave functions is given for some levels of $^{73-81}$Ge isotopes together with yrast levels in Table~3. In Table
sum of the contributions (intensities) from particle
partitions having contribution greater than 1\% is denoted by $S$, the maximum contribution from a single partition by $M$
and the total number of partitions
contributing to $S$ by $N$. The deviation of $S$ from 100\%
is due to high configuration mixing. The increase
in $N$ is also a signature of larger configuration mixing. The extent of configuration mixing is high in $^{73}$Ge and low in $^{79}$Ge. 
Indeed, the extent of configuration mixing is high in the isotopes far from the closed shell. All the three interactions 
predict $\nu(g_{9/2}^{9})$ configuration with probability 42\% (JUN45), 36\%(jj44b) and 80\%($fpg$).

\section{\label{sec4}Electromagnetic properties}
The calculated $B(E2)$ transition probabilities are given in Table~4. For this  effective charges $e_p$=1.5, $e_n$=0.5 
are used in the calculation. Four experimental data for $^{73}$Ge and one for $^{75}$Ge and $^{81}$Ge each are available.  Reasonable result is predicted for $5/2_1^+\rightarrow9/2_1^+$ transition  by all three interaction for $^{73}$Ge.
For the other $E2$ transitions in  $^{73}$Ge the values predicted by JUN45 and jj44b are less than in the experiment, while $fpg$ results are in 
good agreement with the experimental data. The $fpg$ interaction results are reasonable even with above effective charges because now the model space is enlarged,
i.e. we have included $f_{7/2}$ orbital.
The results of quadrupole moments with the three different interactions are shown in Table~\ref{t_q}.
The calculated quadrupole moments by $fpg$ are in good agreement with the existing experimental data.  This
shows the importance of $\pi f_{7/2}$ orbital, which was proposed  in~\cite{Cheal10}, though all calculations predict correct 
sign of the experimental data.
For the $^{75}$Ge, JUN45 predicts
for all the levels negative value of the quadrupole moments, while jj44b predicts all positive values. The $fpg$ calculation gives positive
quadrupole moment for $9/2^+$ and negative quadrupole moment for $7/2^+$ and $5/2^+$. For $^{77}$Ge only JUN45 predicts negative value for quadrupole
moments and the other two predicts positive value. The experimental data for magnetic moments are available only for $^{73}$Ge. In Table~\ref{t_mm},
we compare calculated values of magnetic moments with the experimental data for $^{73}$Ge. We also have presented predicted values of magnetic moments by the three interactions for the remained isotopes considered.
It is seen from Table~\ref{t_mm} that for the $^{73}$Ge calculated values of magnetic moments are in better agreement with the experiment when JUN45 interaction is used.  From Tables 4-6 we can see that the transition rates are strongly enhanced while the quadrupole moments are of the order of the single-particle ones. The large $B(E2)$ values are due to strong dynamical collectivity as even-even nuclei of this region lose the magicity properties.  This is
also seen from Fig. 1.

\section{\label{sec5}Conclusions}

In the present work, the results of large-scale
shell-model calculations are reported for neutron-rich
even--odd isotopes of Ge with $A=73-81$ in two spaces: full $f_{5/2}pg_{9/2}$ space
and $fpg_{9/2}$ space with $^{48}$Ca core.
It was shown that better results are obtained with jj44b interaction for the nuclei further from
closed shell. The results for all calculations become improved reaching to $N\sim50$ closed shell, especially for predicting ground
states. The $E2$ transitions, quadrupole moments and magnetic moments analysis  show the importance of proton excitations across $Z=28$ shell for $fpg_{9/2}$ space. 
Recently modern effective interaction in  $fpg_{9/2}d_{5/2}$ space \cite{Lenzi09} have been developed by Strasbourg-Madrid 
group to study collective phenomena in $fp$ shell nuclei. It remains to be seen if this interaction can describe the experimental data for $Z>28$.\\

This work was supported in part by Conacyt, M\'exico, and by DGAPA, UNAM project IN103212.  MJE acknowledges support
from grant No. 17901 of CONACyT projects CB2010/155633 and from grant F2-FA-F177 of Uzbekistan
Academy of Sciences.

\newpage
%\begin{center}
%{\bf Caption for Figures:}
%\end{center}

{\bf Fig. 1. } Experimental $E(2_1^+)$ for even--even nuclei.

\vspace{4mm}

{\bf Fig. 2. }   Effective single-particle energies of proton orbits for Cu 
isotopes for JUN45 and jj44b interaction.

\vspace{4mm}

{\bf Fig. 3. }  Experimental data \cite{nndc}  for $^{73}$Ge compared
with the results of large-scale shell-model calculations using three different effective interactions.

\vspace{4mm}

{\bf Fig. 4. } The same as in Fig. 3, but for $^{75}$Ge.

\vspace{4mm}

{\bf Fig. 5. }   Experimental data \cite{kay09} for $^{77}$Ge compared
with the results of large-scale shell-model calculations using three different effective interactions.

\vspace{4mm}

{\bf Fig. 6. }  The same as in Fig. 3, but for $^{79}$Ge.

\vspace{4mm}

{\bf Fig. 7. }  The same as in Fig. 3, but for $^{81}$Ge.

\newpage

\newpage

\newpage

\newpage

\newpage

\newpage

\newpage

\newpage
\begin{table}
\caption{ Single-particle energies for both protons and neutrons (in MeV) for JUN45, jj44b, and $fpg$ interactions.}
\label{t_spe}
\begin{center}
%\resizebox{!}{4.5cm}{
\begin{tabular}{c|c|c|c|c}
\hline
Orbital  & ~~JUN45 &~~ jj44b & Orbital&~~ $fpg$ \\		   
\hline
 1$p_{3/2}$&~~ -9.8280  & -9.6566     &~~ 0$f_{7/2}$    &~~ 0.000   \\
\hline
 0$f_{5/2}$&~~ -8.7087 &~~ -9.2859   &~~ 1$p_{3/2}$ &~~ 2.000   \\
\hline
 1$p_{3/2}$&~~-7.8388  &~~ -8.2695    &~~   0$f_{5/2}$  &~~ 6.500   \\
\hline
 0$g_{9/2}$&~~-6.2617  &~~ -5.8944   &~~   1$p_{3/2}$ &~~ 4.000   \\
\hline
           &~~         &~~            &~~ 0$g_{9/2}$   &~~ 9.000   \\
\hline            
\end{tabular}
\end{center}
\end{table} 
  
 \newpage    
\begin{table}[h]
\caption{ Dimensions of the shell-model matrices for  $^{73-79}$Ge odd isotopes
in the $m$-scheme for $f_{5/2}pg_{9/2}$ and $fpg_{9/2}$ spaces ( Note that for $fpg_{9/2}$ space we allowed 
truncation as explained in the text)}
\label{t_dim}
\begin{center}
\resizebox{16.4cm}{!}{ 
\begin{tabular}{c|c|c|c|c|c|c|c|c}
\hline
 \multicolumn{5}{c|}{${f_{5/2}pg_{9/2}}$  space($^{56}$Ni core)}   & \multicolumn{4}{|c}{$fpg_{9/2}$ space($^{48}$Ca core)}\cr
\hline
% &  \multicolumn{2}{c|} \\ 
 $J^\pi$ &~$^{73}$Ge &~~~~~ $^{75}$Ge&~~~~~ $^{77}$Ge&~~~~ $^{79}$Ge&~~~~ $^{73}$Ge&~~~~~ $^{75}$Ge&~~~~~ $^{77}$Ge&~~~~ $^{79}$Ge\cr
\hline
1/2$^+$ &~ 108292130& ~ 38369999&~ 6287004& ~ 404427 &~ 20728481& ~36145446& ~ 18495339 &~  2498759  \cr
\hline
3/2$^+$ &~ 105976096& ~  37491875&~ 6126087& ~ 392187 &~ 20215135& ~35320554& ~ 1843772& ~ 2422738  \cr
\hline
5/2$^+$ &~ 101484694& ~ 35793509&~ 5815605& ~ 368609 &~ 19224163& ~33724105& ~ 17171633& ~ 2277123  \cr
\hline
7/2$^+$ &~ 90585177& ~ 33384092 &~~ 5376962& ~  335428  &~ 17822773& ~31456964& ~ 15937301& ~ 2073809  \cr
\hline
9/2$^+$ &~ 87148373& ~ 30413016&~~ 4840125& ~  295256 &~ 16103247& ~ 28658312& ~ 14421188& ~ 1828986  \cr
\hline
11/2$^+$&~ 78112087& ~ 20754112&~~ 4239812 & ~ 251103 &~  14172417& ~25490972& ~ 12716597& ~ 1560718  \cr
\hline
13/2$^+$&~~ 68446203& ~ 23490606&~~3612831&  ~~206288 &~~ 12142557& ~ 22127977& ~ 10921917& ~ 1287454 \cr 
\hline
1/2$^-$ &~ 108295871 & ~ 38365520&~ 6291153 & ~ 405940 &~ 22834192 & ~31975232& ~ 16712921&~  2968335  \cr
\hline
3/2$^-$ &~ 105977901 & ~ 37489995&~ 6128111& ~ 392899 &~  22276435& ~31222237& ~ 16302500& ~ 2883213  \cr
\hline
5/2$^-$ &~ 101483728& ~ 35795378&~ 5814563& ~ 368174 &~ 21199168& ~29766124& ~ 15510223& ~ 2720053 \cr
\hline
7/2$^-$ &~ 90582567& ~ 33388560&~~ 5373683 & ~ 334156 &~ 19674754& ~27701434& ~ 14390050& ~ 2492022 \cr
\hline
9/2$^-$ &~ 87145850 & ~ 30417768&~~ 4836489& ~293850 &~ 17802156 & ~ 25157862& ~13015630& ~ 2216668 \cr
\hline
11/2$^-$&~ 68447806& ~ 20756608&~~ 4237803& ~ 250276 &~ 15696746& ~ 22287314& ~11472529& ~ 1913639  \cr
\hline
13/2$^-$&~~68446203& ~  23490107&~~ 3613192 &~206306&~~ 13479278& ~ 19249807& ~ 9849959 & ~ 1602512 \cr 
\hline

\end{tabular}}
\end{center}
\end{table}

\newpage
\begin{table} 
 \caption{\label{conf} 
The extent of configuration mixing involved in $^{73-81}$Ge odd
isotopes for the different states (for each state the numbers quoted are $S$, 
sum  of the contributions  from  particle
partitions,  each  of  which  is contributing greater than 1\%;
 $M$, maximum contribution from  a  single  partition; and
$N$, total number of partitions contributing to $S$ ) (Note
that $S$ and $M$ are in percentage)} 
\begin{center}
\resizebox{16.4cm}{!}{ %\resizebox{!}{7.5cm}{
\begin{tabular}{c|c|c|c|c|c|c|c|c|c|c|c|c|c|c|c|c|c|c|c|c} 
\hline
% $J^\pi$&$^{61}_{27}$Co$_{34}$
$J^\pi$& \multicolumn{6}{|c|}{$^{73}_{32}$Ge$_{41}$}& $J^\pi$
&\multicolumn{6}{|c|}{$^{75}_{32}$Ge$_{43}$}&$J^\pi$&\multicolumn{6}{c}{$^{77}_{32}$Ge$_{45}$}\cr
\cline{2-7}\cline{9-14}\cline{16-21}
&\multicolumn{3}{|c|}{JUN45}&\multicolumn{3}{|c|}{jj44b}&&\multicolumn{3}{|c|}{JUN45}&\multicolumn{3}{|c|}{jj44b}&
&\multicolumn{3}{c|}{JUN45}&\multicolumn{3}{|c}{jj44b}\cr
\cline{2-7}\cline{9-14}\cline{16-21}
       &$S$&~ $M$&~ $N$    &$S$&~ $M$&~ $N$&   &$S$&~ $M$&~ $N$          &$S$&~
$M$&~ $N$  &       &$S$~ &$M$~ &$N$     &$S$~ &$M$~ &$N$    \cr
%\cline{2-4}\cline{6-8}\cline{10-12}

\hline
$9/2_1^+$  & 51.3 & ~12.2 & ~17  &~50.7 & ~9.2& ~18  & $9/2_1^+$ &~57.3 & ~14.8 &
~17  &~56.7 & ~17.9& ~19  & $9/2_1^+$  &69.8 & ~19.5 & ~21  & ~67.6 & ~18.7 & ~21\cr
\hline
$7/2_1^+$  & 59.2 & ~16.6 & ~20  &~53.3 & ~9.2& ~19  &$7/2_1^+$  &~60.0 & ~15.7 &
~18  &~56.8 & ~18.4& ~19      &$7/2_1^+$   & 69.7& ~17.6&~23  & ~69.4 & ~20.6 &
~20\cr
\hline
$5/2_1^+$  & 49.7 & ~8.2 & ~19  &~61.4 & ~11.6& ~20 & $5/2_1^+$  &~64.4 & ~16.6 &
~16  &~65.8 & ~23.6& ~20     & $5/2_1^+$   & 71.5& ~20.0& ~21 & ~72.0 & ~ 9.7 &
~24\cr
\hline
$1/2_1^-$  & 63.8& ~21.7 & ~20  &~59.6 & ~20.5& ~17 & $1/2_1^-$   &~70.2 & ~20.5 &
~22  &~63.9 & ~20.7& ~16      & $1/2_1^-$    & 76.8& ~29.4& ~18 & ~71.9 & ~14.2 &
~22\cr
\hline
$5/2_1^-$  & 54.6& ~11.6 & ~18  &~56.4& ~9.9& ~21 &$5/2_1^-$     &~67.3 & ~23.3 &
~20  &~61.0 & ~8.9& ~22       &$5/2_1^-$    & 78.6& ~32.9& ~19& ~74.5 & ~28.9 &
~19\cr
\hline
$3/2_1^-$  & 57.2 & ~17.4 & ~16  &~56.3 & ~16.3& ~17 &$3/2_1^-$    &~62.6 & ~17.2 &
~20  &~65.9 & ~18.4& ~19        &$3/2_1^-$    &70.3& ~22.5& ~18& ~69.7 & ~9.2 &
~23\cr
\hline
$J^\pi$& \multicolumn{6}{|c|}{$^{79}_{32}$Ge$_{47}$}& $J^\pi$
&\multicolumn{6}{|c|}{$^{81}_{32}$Ge$_{49}$} \cr
\cline{2-7}\cline{9-14}
&\multicolumn{3}{|c|}{JUN45}&\multicolumn{3}{|c|}{jj44b}&&\multicolumn{3}{|c|}{JUN45}&\multicolumn{3}{|c|}{jj44b}\cr
\cline{2-7}\cline{9-14}
       &$S$&~ $M$&~ $N$    &$S$&~ $M$&~ $N$&   &$S$&~ $M$&~ $N$          &$S$&~
$M$&~ $N$     \cr
\cline{1-14}
$9/2_1^+$  & 85.7 & ~34.4 & ~18  &~80.8 & ~28.5& ~18  & $9/2_1^+$ &~97.7 & ~31.8 &
~11  & ~97.5 & ~36.4 & ~12  \cr
\cline{1-14}
$7/2_1^+$  & 86.9 & ~38.7 & ~14  &~86.7 & ~35.2& ~15  &$7/2_1^+$  &~96.8& ~52.9&~9 
& ~97.6 & ~42.8 & ~11   \cr 
\cline{1-14}
$5/2_1^+$   & 84.3 & ~30.3 & ~17  &~82.9 & ~25.7& ~20& $5/2_1^+$  &~96.4& ~48.7& ~9
& ~96.1 & ~40.0 & ~ 9    \cr
\cline{1-14}
$1/2_1^-$   & 85.3 & ~27.1 & ~17  &~84.5 & ~27.7& ~17& $1/2_1^-$   &~96.1& ~48.3& ~6
& ~93.1 & ~54.9 & ~11 \cr     
\cline{1-14}
$5/2_1^-$ & 84.4 & ~20.8 & ~17  &~ 87.1& ~33.4& ~14&$5/2_1^-$     &~96.5& ~48.7& ~9&
~90.3 & ~21.8 & ~13     \cr 
\cline{1-14}
$3/2_1^-$  & 82.2 & ~20.9 & ~ 16 &~83.1 & ~19.4& ~17&$3/2_1^-$    &~95.4& ~38.2& ~7&
~92.1 & ~33.6 & ~13   \cr     
\cline{1-14}
\end{tabular}}
\end{center}
\end{table}

\newpage
\begin{table}
\caption{ Calculated and experimental $B(E2)$ values for some transitions with the three different interactions (the effective charges $e_p$=1.5, $e_n$=0.5 
are used in the calculation)}
\label{t_be2}
\begin{center}
%\resizebox{7.5cm}{4.5cm}{
\begin{tabular}{c|c|c|c|c|c|c}
\hline
         &            &                    & \multicolumn{4}{c}{$B(E2)$ (W.u.)} \\
\cline{4-7}
Nucleus &  $I_i^\pi$  $\rightarrow$ $I_i^\pi$ &  $E_\gamma$ (keV)  & Experiment &~~JUN45 &~~ jj44b &~~ $fpg$ \\		   
\hline
 $^{73}$Ge  & 5/2$_1^+$  $\rightarrow$ 9/2$_1^+$    &  13   &  23.1(8) &~~17.18&~~17.38  &~14.50 \\
	    & 7/2$_1^+$  $\rightarrow$ 9/2$_1^+$    &  69    &  41(8) &~~18.38&~~28.61  &~48.72 \\
            & 7/2$_2^+$  $\rightarrow$ 5/2$_1^+$    &  486  &  72(21) &~~2.23&~~5.44 &~23.81 \\
            & 7/2$_2^+$  $\rightarrow$ 9/2$_1^+$    &  499  &  6.3(4) &~~2.47&~~0.47&~4.71 \\
\hline
$^{75}$Ge   & 5/2$_1^+$  $\rightarrow$ 7/2$_1^+$    &  52  &  31$^{+24}_{-28}$&~~ 21.45&~~25.70 &~50.48 \\
            & 9/2$_1^+$  $\rightarrow$ 5/2$_1^+$    &  8    & -  &~~1.42&~~2.66  &~13.83 \\
            & 9/2$_1^+$  $\rightarrow$ 7/2$_1^+$    &  60  & -  &~~19.46&~~22.47 &~ 37.51\\
\hline
$^{77}$Ge  & 9/2$_1^+$  $\rightarrow$ 7/2$_1^+$    & 225  & - &~~17.71 &~~21.43 &~ 28.39\\
           & 5/2$_1^+$  $\rightarrow$ 9/2$_1^+$    & 196  & -  &~~3.96&~~6.86  &~35.93 \\
           & 5/2$_1^+$  $\rightarrow$ 7/2$_1^+$    & 421  & -  &~~12.90&~~ 5.85&~ 10.30\\
\hline
$^{79}$Ge  & 9/2$_1^+$  $\rightarrow$ 7/2$_1^+$    & 204.6  & - &~~11.37 &~~11.94 &~ 11.80\\
           & 9/2$_1^+$  $\rightarrow$ 5/2$_1^+$    & -      & - &~~3.79 &~~3.76 &~ 9.53\\
\hline
$^{81}$Ge  & 5/2$_1^+$  $\rightarrow$ 9/2$_1^+$    & 711.2  & 0.0383(20)&~~ 0.83&~~3.07 &~0.52 \\
           & 5/2$_1^+$  $\rightarrow$ 1/2$_1^+$    &32.10   & -  &~~0.06&~~ 0.33 &~2.05 \\
\hline            
\end{tabular}
\end{center}
\end{table} 

\newpage
\begin{table}
\caption{ Electric quadrupole moments, $Q_s$ (in $e$b), with the three different interactions (the effective charges $e_p$=1.5, $e_n$=0.5 are used in the calculation)}
\label{t_q}
\begin{center}
%\resizebox{7.5cm}{4.5cm}{
\begin{tabular}{c|c|c|c|c|c}
\hline
Nucleus & $J^{\pi}$  &$Q_{s,exp}$ &~$Q_{s,JUN45}$  &~~ $Q_{s,jj44b}$  &~$Q_{s,fpg}$  \\		   
\hline
 $^{73}$Ge  &  9/2$^+$    &~~ -0.173(26)   &~~-0.187              &~~ -0.090           &~ -0.293\\
             &  7/2$^+$     &~~ -         &~~-0.439          &~~  -0.144            &~ -0.693 \\
            &  5/2$^+$     &~~ +0.70(8)     &~~+0.114          &~~  +0.374            &~+0.104 \\
\hline
 $^{75}$Ge  &  9/2$^+$    &~~   -  &~~ -0.031            &~~ +0.008              &~  +0.060\\
                      &  7/2$^+$    &~~  -   &~~ -0.089            &~~ +0.087              &~   -0.067\\
                      &  5/2$^+$    &~~  -   &~~ -0.295            &~~ +0.271              &~   -0.539\\
\hline
 $^{77}$Ge  &  9/2$^+$    &~~ -     &~~ +0.096         &~~ +0.176          &   ~ +0.272\\
                      &  7/2$^+$    &~~  -   &~~  +0.237         &~~ +0.470          &   ~+0.571\\
                      &  5/2$^+$    &~~  -   &~~  -0.179         &~~ -0.137          &   ~-0.054\\
\hline                      
$^{79}$Ge  &  9/2$^+$    &~~ -     &~~+0.219         &~~ +0.285           &   ~ +0.464\\
                     &  7/2$^+$    &~~ -    &~~+0.415            &~~ +0.446          &~+0.539\\
                      &  5/2$^+$    &~~ -    &~~  -0.048        &~~ +0.012          &   ~+0.297\\
\hline                     
$^{81}$Ge  &  9/2$^+$    &~~ -     &~~+0.374          &~~+0.452           &   ~ +0.367\\  
                     &  7/2$^+$    &~~ -    &~~+0.076            &~~ +0.098         &~+0.173\\
                     &  5/2$^+$    &~~  -   &~~  +0.313        &~~ +0.339          &   ~+0.056\\               
\hline                             
\end{tabular}
\end{center}
\end{table}

\newpage
\begin{table}
\caption{ Magnetic moments, $\mu$ (in $\mu_N$), with  three different interactions (for $g_{s}^{eff}$ = 0.7$g_{s}^{free}$ )}
\label{t_mm}
\begin{center}
%\resizebox{7.5cm}{4.5cm}{
\begin{tabular}{c|c|c|c|c|c}
\hline
Nucleus & $J^{\pi}$  &$\mu_{exp}$ &~$\mu_{JUN45}$  &~~ $\mu_{jj44b}$ &~$\mu_{fpg}$  \\		   
\hline
 $^{73}$Ge  &  9/2$^+$    &~~ -0.879(2)   &~~-0.961              &~~ -0.780         &~ -0.496\\
             &  7/2$^+$     &~~ -         &~~-0.826         &~~  -0.766            &~ -0.646 \\
            &  5/2$^+$     &~~ -1.08(3)     &~~-1.361          &~~  -0.767           &~-1.198 \\
\hline
 $^{75}$Ge  &  9/2$^+$    &~~   -  &~~ -0.940            &~~ -0.823              &~  -0.552\\
                      &  7/2$^+$    &~~  -   &~~ -0.885           &~~ -0.850              &~   -0.519\\
                      &  5/2$^+$    &~~  -   &~~ -0.667            &~~-0.626             &~   -0.503\\
\hline
 $^{77}$Ge  &  9/2$^+$    &~~ -     &~~ -0.954         &~~ -0.781          &   ~ -0.728\\
                      &  7/2$^+$    &~~  -   &~~  -0.939         &~~ -0.865          &   ~-0.791\\
                      &  5/2$^+$    &~~  -   &~~  -0.912         &~~ -0.947          &   ~-1.019\\
\hline                      
$^{79}$Ge  &  9/2$^+$    &~~ -     &~~-0.978        &~~ -0.938           &   ~ -0.947\\
                     &  7/2$^+$    &~~ -    &~~-0.854            &~~ -0.831          &~-0.826\\
                      &  5/2$^+$    &~~ -    &~~  -1.511        &~~ -1.483          &   ~-1.309\\
\hline                     
$^{81}$Ge  &  9/2$^+$    &~~ -     &~~-1.014         &~~-1.012           &   ~ -0.947\\  
                     &  7/2$^+$    &~~ -    &~~-0.362            &~~ -0.645        &~-0.268\\
                     &  5/2$^+$    &~~  -   &~~  -1.851       &~~ -2.411          &   ~-2.156\\               
\hline                             
\end{tabular}
\end{center}
\end{table}


\begin{thebibliography}{99}
\bibliographystyle{maik}

\bibitem{Rother11}
W. Rother {\it et al.},
Phys.\ Rev.\ Lett. {\bf106}, 022502 (2011).

\bibitem{Aoi09}
N. Aoi {\it et al.},
Phys.\ Rev.\ Lett. {\bf102}, 012502 (2009).

\bibitem{Recchia12}
F.~Recchia {\it et al.},
Phys.\ Rev.\ C {\bf85}, 064305 (2012).


\bibitem{Kaneko08}
K.~Kaneko, Y. Sun, M.~Hasegawa, and T.~Mizusaki,
Phys.\ Rev.\ C {\bf78}, 064312 (2008).

\bibitem{Sri09}
P. C.~Srivastava and I.~Mehrotra,
J.\ Phys.\ G\ {\bf36},  105106 (2009).

\bibitem{Sria}
P. C.~Srivastava and I.~Mehrotra, 
Phys.\ Atom.\ Nucl.\ {\bf73}, 1656 (2010)[Yad.\ Fiz.\ {\bf73}, 1703 (2010)].

\bibitem{Srie}
P. C.~Srivastava and I.~Mehrotra, 
Eur.\ Phys. \ J. A \ {\bf45}, 185 (2010).

\bibitem{Sriaa}
P. C.~Srivastava and V. K.B.~Kota, 
Phys.\ Atom.\ Nucl.\ {\bf74}, 971 (2011)[Yad.\ Fiz.\ {\bf74}, 1000 (2011)].

\bibitem{Lenzi09}
S. M.~Lenzi, F. Nowacki, A. Poves, and K. Sieja, Phys.\ Rev.\ C {\bf82}, 054301 (2010).

\bibitem{Sun12}
Y. Sun, Y.C. Yang, H. Jin, K. Kaneko, and S. Tazaki,
Phys.\ Rev.\ C {\bf85}, 054307 (2012). 

\bibitem{Cheal10}
B. Cheal {\it et al.},
Phys.\ Rev.\ Lett. {\bf104}, 252502 (2010).


\bibitem{Rodal05}
E. Padilla-Rodal {\it et al.}, 
Phys.\ Rev.\ Lett. {\bf94} 122501 (2005).

\bibitem{Zamick11} 
S. J.Q. Robinson, L. Zamick, and Y. Y. Sharon,
Phys.\ Rev.\ C {\bf83} 027302  (2011).

\bibitem{Sri12con}
J. G. Hirsch and P. C. Srivastava,
J. Phys.: Conf. Ser. {\bf387} 012020 (2012).


\bibitem{Gade10} 
A. Gade {\it et al.}, 
Phys.\ Rev.\ C {\bf81} 064326  (2010).

\bibitem{yosinaga}
N. Yoshinaga, K. Higashiyama, and P.H. Regan, 
Phys.\ Rev.\ C {\bf78},  044320 (2008).


\bibitem{Honma09}
M. Honma, T. Otsuka, T. Mizusaki and M. Hjorth-Jensen, 
Phys.\ Rev.\ C {\bf80},  064323 (2009).

\bibitem{pcs_ga}
P. C.  Srivastava,
J.\ Phys.\ G\ {\bf39},  015102 (2012).

\bibitem{brown}
B.A. Brown and A.F. Lisetskiy (unpublished).

\bibitem{kum12} 
G. J. Kumbartzki {\it et al.}, 
Phys.\ Rev.\ C {\bf85} 044322  (2012).

\bibitem{saha12} 
S. Saha {\it et al.}, 
Phys.\ Rev.\ C {\bf86} 034315  (2012).

\bibitem{Sorlin02}
O. Sorlin {\it et al.}, 
Phys.\ Rev.\ Lett. {\bf88}, 092501 (2002).

\bibitem{Pov01}
A.~Poves, J. Sanchez-~Solano, E.~Caurier, and F. Nowacki,
Nucl.\ Phys. A {\bf 694},  157 (2001).

\bibitem{Nowacki96}
F.~Nowacki, Ph.D. Thesis
(IRes, Strasbourg, 1996).

\bibitem{Kahana69}
S.~Kahana, H.C.~Lee, and C.K.~Scott,
Phys.\ Rev.\  {\bf180},  956 (1969).


\bibitem{Antoine}
E. Caurier, G. Mart\'inez-Pinedo, F. Nowacki {\it et al.},
Rev.\ Mod.\ Phys. {\bf77}, 427 (2005).


\bibitem{hak08}
J. Hakala {\it et al.}, Phys.\ Rev.\ Lett. { \bf101}, 052502 (2008).

\bibitem{iwa08}
H. Iwasaki {\it et al.}, Phys.\ Rev. C {\bf 78}, 021304(R) (2008).

\bibitem{nndc} www.nndc.bnl.gov.

\bibitem{kay09}
B. P.~Kay {\it et al.},
Phys.\ Rev.\ C {\bf80},  017301 (2009).

\end{thebibliography}
\end{document}